\documentclass[twocolumn,prb,amsmath,amssymb,nopacs]{revtex4}
\usepackage{graphicx}
\usepackage{dcolumn}
\usepackage{bm}
\usepackage{latexsym}
\usepackage{amstext}
\usepackage{amsxtra}
\usepackage{color}

\newcommand{\ket}[1]{|#1\rangle}
\newcommand{\bra}[1]{\langle#1|}
\newcommand{\bs}[1]{\boldsymbol{#1}}

\usepackage{times}

\begin{document}

\title{Coherent dipole-dipole coupling between two single atoms at a F\"orster resonance}

\author{S. Ravets, H. Labuhn, D. Barredo, L. B\'eguin, T. Lahaye, and A. Browaeys}
\affiliation{Laboratoire Charles Fabry, UMR 8501, Institut d'Optique, CNRS, Univ Paris Sud 11, \\
2 avenue Augustin Fresnel, 91127 Palaiseau Cedex, France}

\maketitle

{\bf Resonant energy transfers, i.e. the non-radiative redistribution of an electronic excitation between two particles coupled by the dipole-dipole interaction, lie at the heart of a variety of chemical and biological phenomena~\cite{ret}, most notably photosynthesis. In 1948, F\"orster established the theoretical basis of fluorescence resonant energy transfer (FRET) \cite{Forster1948}, paving the ground towards the widespread use of FRET as a ``spectroscopic ruler'' for the determination of nanometer-scale distances in biomolecules~\cite{Stryer1967}. The underlying mechanism is a coherent dipole-dipole coupling between particles, as already recognized in the early days of quantum mechanics~\cite{Clegg2006}, but this coherence was not directly observed so far. Here, we study, both spectroscopically and in the time domain, the coherent, dipolar-induced exchange of electronic excitations between two single Rydberg atoms separated by a controlled distance as large as $15\;\mu{\rm m}$, and brought into resonance by applying a small electric field. The coherent oscillation of the system between two degenerate pair states occurs at a frequency that scales as the inverse third power of the distance, the hallmark of dipole-dipole interactions~\cite{Walker2005}. Our results not only demonstrate, at the most fundamental level of two atoms, the basic mechanism underlying FRET, but also open exciting prospects for active tuning of strong, coherent interactions in quantum many-body systems. }

\begin{figure}[b!]
\centering
\includegraphics[width=\linewidth]{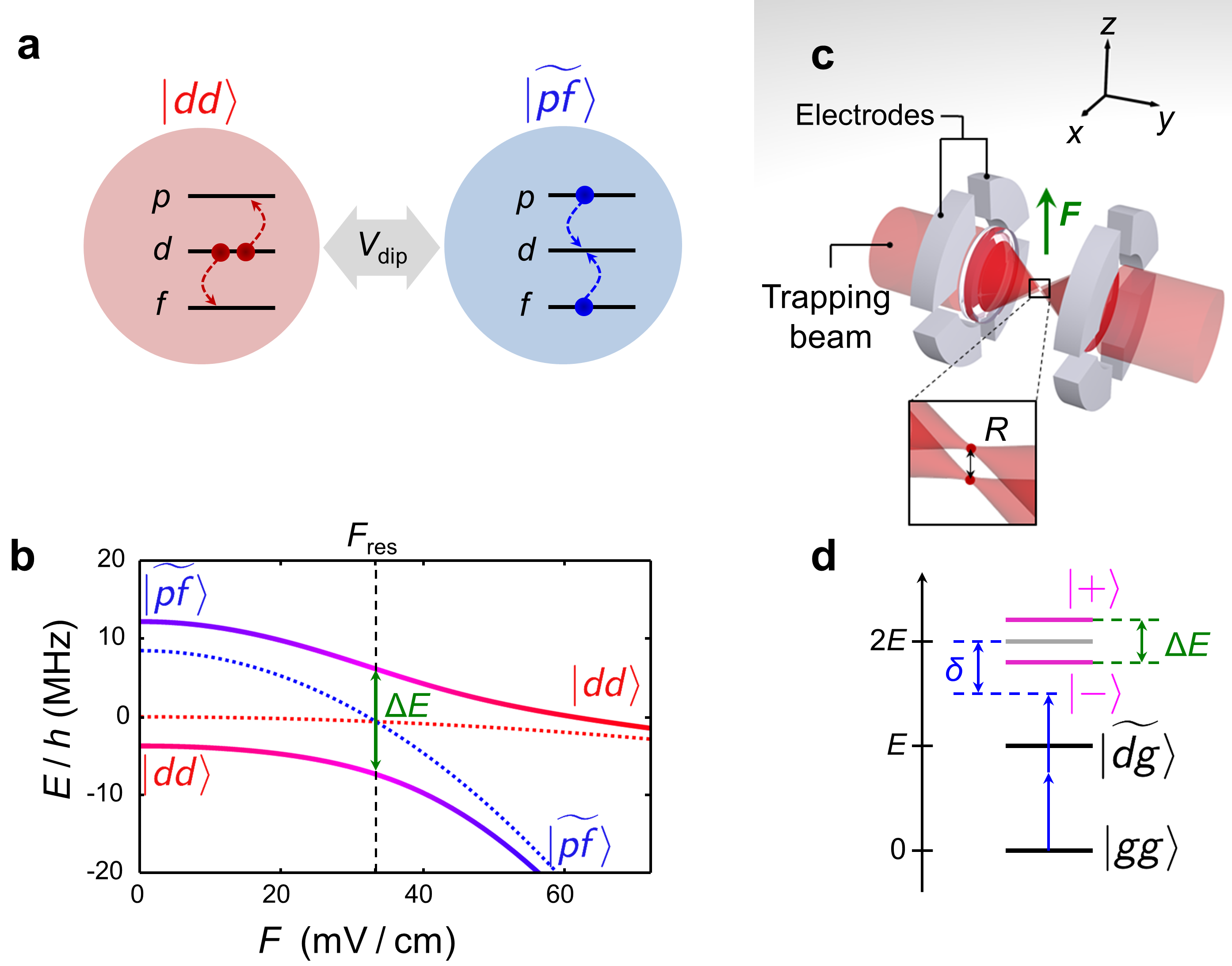}
\caption{{\bf Principle of the experiment.} {\bf a}, The pair states $\ket{dd}$ and $\ket{\widetilde{pf}}$ are coupled by the dipolar interaction at a F\"orster resonance. {\bf b}, Stark effect of the pair states. In the absence of coupling, the two pair-state energy levels (dotted lines) vary when an external electric field $F$ is applied and cross for $F=F_{\rm res}$. The coupled system (solid lines) undergoes an avoided crossing at resonance (vertical dashed line). {\bf c}, Experimental setup. Two individual $^{87}{\rm Rb}$ atoms  are trapped in microscopic optical tweezers separated by a distance $R$. Eight independent electrodes allow to apply a tunable electric field $\bs{F}$ along the internuclear axis $z$. {\bf d}, Structure of the coupled levels at resonance. We scan the excitation laser (blue arrow) to perform a two-atom excitation from $\ket{gg}$ to $\ket{dd}$ via the intermediate, one-excitation state $\ket{\widetilde{dg}}=(\ket{dg}+\ket{gd})/\sqrt{2}$. The detuning between the laser frequency and the position of $\ket{dd}$ in the absence of interactions (dashed horizontal line) is denoted by~$\delta$.}
\end{figure}

The possibility to tune at will coherent interactions in many-body systems by changing an external parameter is one of the key tools enabling quantum simulation. For instance, in ultracold quantum gases, such tuning can be achieved by magnetically-induced Feshbach resonances~\cite{Inouye1998,rmpfeshbach}. Rydberg atoms are another promising platform for the quantum simulation of complex many-body problems, due to the strong interactions associated with their large principal quantum numbers~\cite{Gallagher2005}. They have proved to be an efficient tool for the characterization of non-radiative exchange of energy in resonant collisional processes\cite{Safinya1981}, the study of collective effects~\cite{Comparat2010} and the engineering of quantum states of matter~\cite{Saffman2010}. The observation of the Rydberg blockade effect between individual atoms~\cite{Urban2009,Gaetan2009}, where the strong interaction between Rydberg states inhibits multiple excitations within a blockade sphere, opens the way towards the development of Rydberg quantum simulators~\cite{Weimer2010}. An appealing tool for all those applications is the possibility to tune the strength of the interactions by applying external electric fields using F\"orster resonances~\cite{Anderson1998,Mourachko1998,Anderson2002,Mudrich2005,Vogt2007,Reinhard2008,Ryabtsev2010,Gunter2013}. So far, due to inhomogeneities in the atomic ensembles used in experiments, only indirect evidence for the coherent character of the interaction could be obtained~\cite{Nipper2012,Nipper2012x}. 

Here, we study a system of two single atoms at a F\"orster resonance. We first perform a spectroscopic measurement of the energies of the two-atom states as a function of the applied electric field, and observe directly the avoided crossing between pair states induced by the dipole-dipole interaction. The splitting at resonance is observed to scale as $1/R^3$ as a function of the distance $R$ between the atoms. In a second experiment, we prepare the system in a given pair state away from resonance, and switch to resonance for a controlled time, revealing the coherent oscillation between the two degenerate pair states induced by the dipole interaction. These results open the way to real-time tuning of interactions for quantum simulation with Rydberg atoms~\cite{Saffman2010, Weimer2010}.

\begin{figure*}[t]
\centering
\includegraphics[width=15cm]{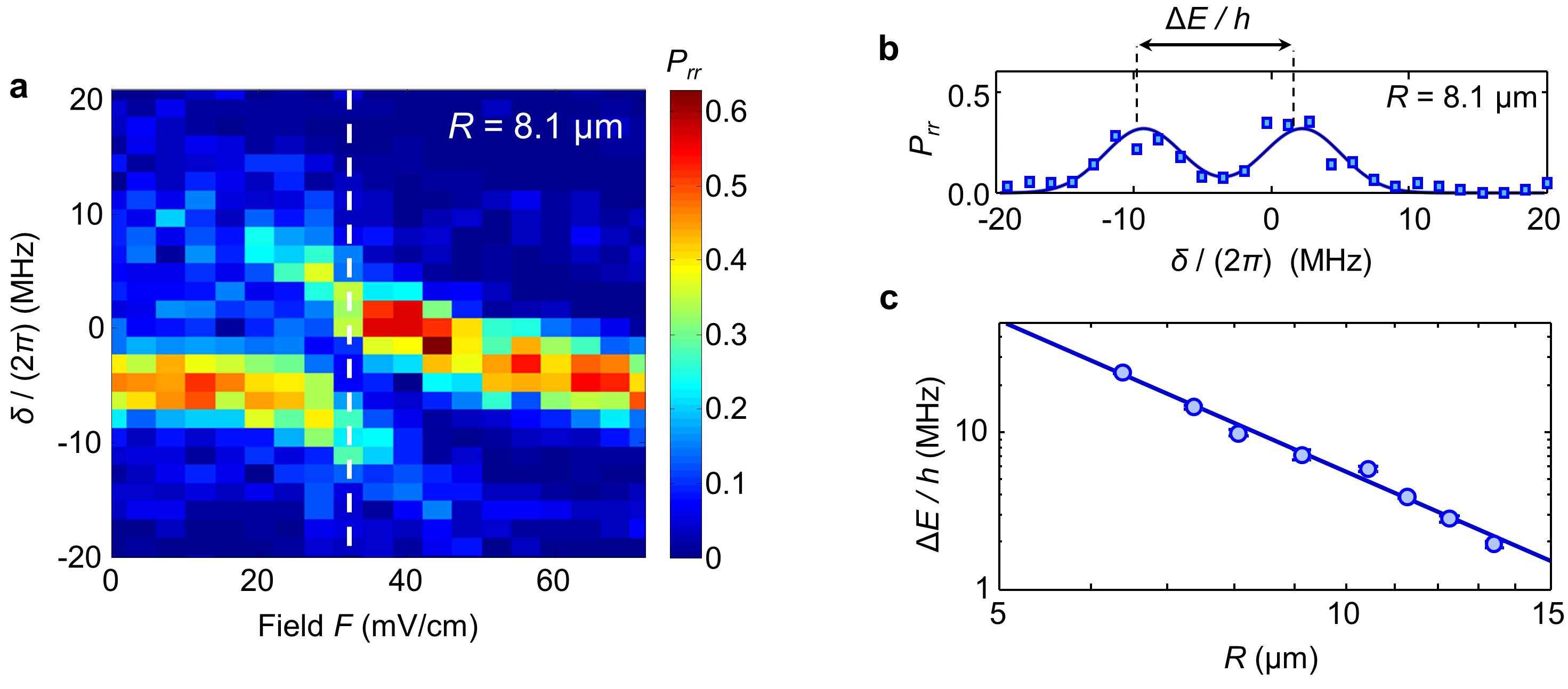}
\caption{{\bf Spectroscopy at the F\"orster resonance.} {\bf a}, Two-atom excitation spectrum where the population of doubly-Rydberg-excited states, $P_{rr}$, is plotted as a function of $\delta$ and $F$, showing the avoided crossing between the resonant pair states. The white dashed line indicates the position of the F\"orster resonance. {\bf b}, Spectrum $P_{rr}(\delta)$ at resonance for $R=8.1\,{\rm \mu m}$. The solid line is a fit by a double gaussian. {\bf c}, Double logarithmic plot of the splitting $\Delta E$ between the peaks, at resonance, as a function of $R$. The solid line shows a fit by a power law, giving an exponent $-3.2 \pm 0.2$.}
\end{figure*}

Two atoms located at positions $\bs{R}_1$ and ${\bs R}_2$ interact through the dipole-dipole interaction described by the hamiltonian 
\begin{equation}
\hat{V}_{\rm dip} =\frac{1}{4\pi \varepsilon_0} \left( \frac{\hat{\bs{\mu}}_1 \cdot \hat{\bs{\mu}}_2-3(\hat{\bs{\mu}}_1 \cdot \bs{n})(\hat{\bs{\mu}}_2 \cdot \bs{n})}{R^3} \right),
\end{equation}
where $\hat{\bs{\mu}}_i$ is the electric dipole moment of atom $i$ ($i=1,2$), $\bs{R}=\bs{R}_2-\bs{R}_1$, and $\bs{n}=\bs{R}/R$. When the two atoms are prepared in the same state, $\hat{V}_{\rm dip}$ has no effect to first order as the average value of the dipole moment vanishes in an atomic eigenstate. Second-order perturbation theory gives rise to an energy shift of the atom pair, which results in the van der Waals interaction~\cite{Beguin2013} $U_{\rm vdW} \propto R^{-6}$. However, resonance effects between two Rydberg atoms can occur when two pair states are energetically degenerate~\cite{Walker2005}, and in this case the dipolar interaction manifests itself at first order. Such a resonance, called a F\"orster resonance in analogy with the FRET mechanism at work in photochemistry, can be achieved using small electric fields to Stark-tune the energy of the pair states. 

In this work, we use the states $\ket{p}=\ket{61P_{1/2},m_J=1/2}$, $\ket{d}=\ket{59D_{3/2},m_J=3/2}$ and $\ket{f}=\ket{57F_{5/2},m_J=5/2}$ of $^{87}{\rm Rb}$ atoms. The pair states $\ket{dd}$, $\ket{pf}$ and $\ket{fp}$ are almost degenerate (see Fig.~1a): their \emph{F\"orster defect} $ \Delta_{0} = (E_{pf} - E_{dd})/h$, in the absence of an electric field, is only $8.5~{\rm MHz}$ ($h$ is Planck's constant). Using the differential Stark effect between $\ket{dd}$ and $\ket{pf}$, they can be brought to exact resonance by applying an electric field $F_{\rm res} \simeq 32~{\rm mV/cm}$ (see Fig.~1b)\footnote{The small electric fields at play ensure that we are working in a regime of induced dipoles, as opposed to the rigid dipoles obtained for larger electric fields.}. At resonance, the eigenstates of the interacting system are $\ket{\pm} = (\ket{dd} \pm \ket{\widetilde{pf}}) / \sqrt{2}$, where $\ket{\widetilde{pf}}=(\ket{pf}+\ket{fp}) / \sqrt{2}$. If the system is initially prepared in $\ket{dd}$, it will thus oscillate between the two degenerate electronic configurations with a frequency given by the dipolar coupling $2\sqrt{2}C_3/R^3$ (where $C_3\sqrt{2}/R^3 = \bra{dd} \hat{V}_{\rm dip} \ket{\widetilde{pf}}$). In particular, after half a period of interaction, the system has evolved to the entangled state $\ket{\widetilde{pf}}$. 

Our experimental setup has been described previously~\cite{Beguin2013,Nogrette2014}. We trap two single laser-cooled atoms in optical tweezers separated by a controlled distance $R$ of a few microns (see Fig. 1c).  A set of eight independent electrodes allows us to apply a controlled electric field $F$ aligned with the internuclear axis~\cite{Low2012}. We couple the ground state $\ket{g}=\ket{5S_{1/2},F=2,m_F=2}$ to the Rydberg state $\ket{d}$ using a two-photon transition with effective Rabi frequency $\Omega$. The readout of the states of the atoms is ensured by shining resonant light at 780~nm on the atoms, giving a fluorescence signal only if the atom is in $\ket{g}$. Repeating the same sequence $\sim 100$ times allows us to reconstruct the populations $P_{ij}$ ($i,j$ take on the values $g,r$, where $r$ stands for any of the Rydberg states $p$, $d$, and $f$).

\begin{figure*}[t]
\centering
\includegraphics[width=15cm]{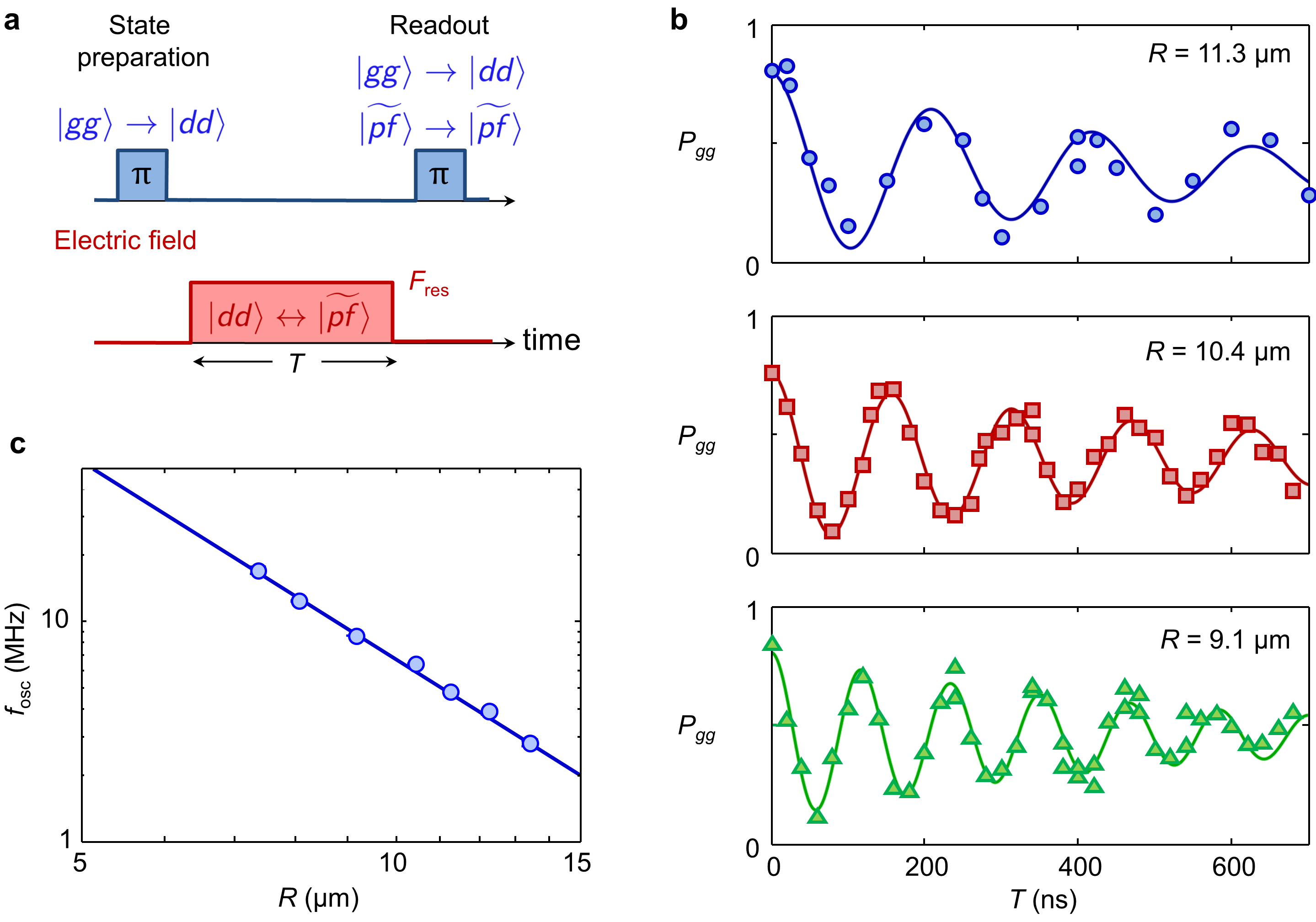}
\caption{{\bf Observation of the oscillation between two atoms.} {\bf a}, Experimental sequence. After exciting the atoms to $\ket{dd}$ (first $\pi$ pulse), we switch on the resonant interaction for a variable amount of time $T$ by tuning the electric field to $F_{\rm res}$. The system oscillates back and forth between the two pair states $\ket{dd}$ and $\ket{\widetilde{pf}}$ for the duration $T$. A deexcitation pulse (second $\pi$ pulse) couples back the $\ket{dd}$ part of the resulting state to $\ket{gg}$. {\bf b}, Evolution of $P_{gg}$ as a function of the interaction time. $P_{gg}$ oscillates at a frequency $f_{\rm osc}$ which depends on $R$. The solid lines are fits by a damped sine wave with frequency $f_{\rm osc}$. {\bf c}, Double logarithmic plot of the oscillation frequency $f_{\rm osc}$ as a function of $R$. The solid line shows a fit by a power law, giving an exponent $-3.0 \pm 0.1$.}
\end{figure*}

We first fix $R=8.1\;{\rm \mu m}$ and perform a spectroscopic measurement to find the position of the F\"orster resonance. Our laser system couples $\ket{gg}$ to $\ket{dd}$ only, and thus partly couples $\ket{gg}$ to the states $\ket{+}$ and $\ket{-}$ at resonance. For increasing values of $F$, we scan the laser detuning $\delta$ (defined with respect to the transition $\ket{gg}$ to $\ket{dd}$, see Fig.1d) and measure the probability $P_{rr}$ for both atoms to be in a Rydberg state (see Fig.~2 a). For $F=0$, we observe a single line centered at $\delta/(2\pi) \approx -5~{\rm MHz}$, corresponding to the attractive van der Waals interaction between the two atoms out of resonance. For $F \approx 20~{\rm mV/cm}$, a repulsive branch appears in the spectrum, a signature of the interaction between $\ket{dd}$ and $\ket{\widetilde{pf}}$. Increasing the field even further allows scanning across the avoided crossing until only one peak is visible again. We reach the F\"orster resonance at the electric field $F_{\rm res}=32\pm4~{\rm mV/cm}$, where we observe two symmetric peaks corresponding to $\ket{\pm}$. 

We then measure the evolution of the spectra at resonance when varying the interatomic distance $R$. To extract the interaction energy in a simple way, we fit the spectra by two Gaussians with a splitting $\Delta E$ between the two peaks (see Fig.2b). When increasing $R$, the splitting between the peaks decreases. Figure~2c shows a double-logarithmic plot of $\Delta E$ versus $R$. The data shows a power-law behavior of exponent $-3.2 \pm 0.2$, consistent with the expected $C_3/R^3$ law. We measure $C_3 = 2.1 \pm 0.1~{\rm GHz\cdot\mu m^3}$, where the error bar is statistical. Systematic effects are an overall $\sim 5\%$ uncertainty in our calibration of $R$, and possible residual light-shifts in the two-photon spectroscopy. Theoretical calculations~\cite{Reinhard2007} give $C_{3,\rm th} \simeq 2.54~{\rm GHz\cdot\mu m^3}$. 

We now study the coherence properties of the system at resonance using a sequence (see Fig.~3a) reminiscent of a pump-probe spectroscopy experiment. We prepare the system in the state $\ket{dd}$ using a $\pi$-pulse.  In order to start in a pure $\ket{dd}$ state, we perform the excitation in the van der Waals regime above resonance ($F \approx 64~{\rm mV/cm}$), where the F\"orster defect is $\Delta(F) \geq h\times 100~{\rm MHz}$ and where interactions are small. We then turn on the resonant interaction for a variable duration $T$, by switching rapidly (risetime below $10$~ns) the field to $F_{\rm res}$. During this time, the two-atom system oscillates between $\ket{dd}$ and the entangled state $\ket{\widetilde{pf}}$, with a frequency $f_{\rm osc}=\Delta E/h= 2\sqrt{2}C_3/(hR^3)$ given by the dipolar coupling. We then apply a deexcitation $\pi$-pulse identical to the first one to read out the state of the system. The deexcitation pulse couples the $\ket{dd}$ component of the system back to $\ket{gg}$. At the end of the sequence we measure the probability $P_{gg}$ to be back in $\ket{gg}$. We observe highly contrasted oscillations between $\ket{dd}$ and $\ket{\widetilde{pf}}$ (see Fig.~3b).

The oscillations between two electronic states are a direct observation of the coherent nature of the coupling underlying F\"orster energy transfer. We fit the oscillation by a damped sine wave to extract the oscillation frequency $f_{\rm osc}$. The damping arises from (i) shot-to-shot fluctuations (on the order of $\pm 200$~nm) in the distance $R$ due to the finite temperature of the atoms in the tweezers, and (ii) electric field noise. Figure~3c shows a double-logarithmic plot of the values of $f_{\rm osc}$ as a function of $R$. The data shows a power-law behavior of exponent $-3.0 \pm 0.1$, again in excellent agreement with the expected $R^{-3}$ behavior. The measured $C_3 = 2.39 \pm 0.03~{\rm GHz\cdot\mu m^3}$ is also close to the theoretical value.

Our results open exciting prospects for real-time tuning of interactions in systems of Rydberg atoms, in particular to switch on and off Ryberg blockade on nanosecond timescales. As an illustration, in the above experiment, when switching $F$ from 64~mV$/$cm (away from resonance) to 32~mV$/$cm (right on resonance), the blockade shift between two atoms separated by $R=10\;\mu{\rm m}$ varies from $U=U_{\rm vdW}\sim h\times 0.2$~MHz (van der Waals regime) up to $U=\Delta E/2\sim h \times 4$~MHz ($C_3/R^3$ regime). If the pair of atoms initially in $\ket{gg}$ is driven with a Rabi frequency $\Omega/(2\pi)\sim1$~MHz, one would observe a strong blockade in the second case, while blockade would be almost totally suppressed in the first situation. This means that, simply by changing the value of the electric field by a few mV$/$cm, we obtain a twenty-fold enhancement of the interaction, and the blockade radius is increased by a factor $\sim 2$ in real time, a feature hard to achieve by other means. 

In conclusion, we directly observed the oscillation between the two degenerate pair states of two single Rydberg atoms at a F\"orster resonance, demonstrating the coherent nature of the mechanism underlying resonant energy transfer. The presence of only two atoms at a controlled distance, with well defined internal states, was fundamental to this study, as disorder in the positions of the atoms would wash out the coherent character of the interaction. A natural extension of this work will consist in measuring the angular dependence of resonant interactions~\cite{Reinhard2007}, in view of tailoring even further the interactions between two particles. Extending our results beyond two particles, to few-body~\cite{Gurian2012} and many-body systems~\cite{Anderson1998,Mourachko1998,Gunter2013}, will enable the study the transport of excitations and the generation of entanglement in fully controlled many-body systems.

{\bf Acknowledgments} We thank C.S. Adams for enlightening discussions, F. Nogrette for technical assistance and M. Besbes for finite-element calculations of the electric field configuration. This work was supported financially by the EU (ERC Stg Grant ARENA, FET-Open Xtrack project HAIRS, EU Marie-Curie program ITN COHERENCE FP7-PEOPLE-2010-ITN-265031 (H.L.)), by the DGA (L.B.), and by R\'egion \^Ile-de-France (LUMAT and Triangle de la Physique, LAGON project). 



\end{document}